\documentclass[aps,prl,twocolumn,groupedaddress]{revtex4}
\usepackage{epsf}
\usepackage{graphicx}

\begin{document}

\title{Charge carriers of different origin in cuprates as revealed by
experiment}
\author{Lev P. Gor'kov}
\affiliation{NHMFL, Florida State University, 1800 E P.Dirac Dr., Tallahassee FL 32310,
USA}
\affiliation{L.D.Landau Institute for Theoretical Physics of the RAS, Chernogolovka
142432, RUSSIA }
\author{Gregory B. Teitel'baum}
\email{grteit@kfti.knc.ru}
\affiliation{E.K.Zavoiskii Institute for Technical Physics of the RAS, Kazan 420029,
RUSSIA }
\date{\today}

\begin{abstract}
The Hall coefficient data for cuprates show that number of carriers exceeds
external doping $x$ at higher $x$ and varies with temperature. Hence, spins
on the Cu-sites are not conserved. Activation energy for thermally excited
carriers equals the energy between the Fermi surface ``arc'' and the band
bottom near the van Hove singularities. Crossover from marginal Fermi
liquid- to pseudogap- regime happens at temperatures at which number of
activated carriers gets comparable with the number of externally doped
holes. Implications for the $(T,x)$-phase diagram of cuprates are discussed.
\end{abstract}

\maketitle



The unifying feature for all cuprates is the presence of one or more CuO$%
_{2} $ --planes. The consensus is that the in-plane electronic constituents,
namely, the Cu $d^{9}$ levels and the oxygen $p$-orbitals determine all
cuprates' physics \cite{1,2}. Even this simplified model turns out to be
difficult for complete theoretical analysis, and properties of cuprates
remain far from being understood. Below, from the experimental stand-point,
we address mainly properties of the single-plane cuprates, La$_{2-x}$Sr$_{x}$%
CuO$_{4}$ (LSCO), the materials best studied by now.

Transition from the Mott insulating state into a metallic and
superconducting (SC) state is driven in cuprates by external doping. We
emphasize from the very beginning that doping, rigorously speaking, is
\textit{not} the thermodynamic path for Metal-Insulator (MI) transition.

It is often assumed that initially holes go onto the oxygen sites, since the
Cu($d^{9}$)- and the oxygen's levels in the parent antiferromagnet (AFM) La$%
_{2}$CuO$_{4}$ are separated by a ``charge transfer gap''\ of order $\sim $%
1.2-1.5 eV \cite{3}. It becomes less obvious with the increase of
concentration, $x$, because charges of dopant ions result in important
changes in the system's energy balance. Indeed, already at rather small
concentrations (as seen from the ARPES data at $x\sim 0.03$ \cite{4}) it is
more proper to resort to bands' description, at least, for the oxygen bands.
This gives rise to the hybridization between oxygen and Cu levels, although
the Cu($d^{9}$) level tends to partially conserve its local character owing
to strong tendency to the Jahn-Teller polaron formation inherent in the $%
d^{9}$- configuration. Occupied neighboring Cu($d^{9}$)-sites experience
strong interactions via mutual local lattice distortions. As to the exchange
spin interactions, it is now clear that they play a secondary role by
coupling spins on the adjacent sites antiferromagnetically \cite{5,6}.

In what follows we discuss some recent experiments that have shed more light
on the problem of the nature and actual number of carriers and on the
stability of the Cu $d^{9}$-hole configuration, the latter being responsible
for existence of local spin $S=1/2$ at a given Cu-site.

There are may be some concerns how to formulate such a question. Indeed, the
($T$,$x$) plane for cuprates, e.g., La$_{2-x}$Sr$_{x}$CuO$_{4}$, is sub-
divided into two main parts by a crossover line, $T^{\ast }(x)$ (see, e.g.,
in \cite{7}). To the right from $T^{\ast }(x)$ lies the so-called marginal
Fermi liquid (MFL) regime \cite{8} that seems to merge with the traditional
FL at even larger $x$. Area on the left hand side is known in the literature
as a ``pseudogap regime''(PG). Although no consensus exists yet regarding
details, the pseudogap regime seems to be spatially inhomogeneous. In
particular, it was suggested (see discussion and references in \cite{9})
that the $T^{\ast }(x)$ is the \textit{line} for start of a 1st order
transition frustrated by the electroneutrality condition in the presence of
rigidly embedded dopants. If so, the area between the $T$-axis $(x=0)$ and $%
T^{\ast }(x)$ could be considered as a miscibility gap for solution of
strontium holes into the La$_{2}$CuO$_{4}$ network, if it were not for the
long range Coulomb forces on the part of rigid Sr$^{2+}$ ions that prohibit
the phase separation on a macroscopic scale. Therefore, at elevated
temperatures this area emerges as the regime of a \textit{dynamical}
competition between two sub-phases (magnetic and metallic), as it was first
suggested in \cite{5}. Such two-component character of the PG side of the
cuprates phase diagram has been firmly established at the analysis of the
NMR experiments in [9]. It turned out that the $^{63}$Cu nuclear relaxation
rate comes about from two independent dissipative processes, related to the
dynamically coexisting phases --islands of incommensurate antiferromagnetic
(ICAF) phase (''stripe phase'' seen in the inelastic neutron's scattering
experiments as IC-peaks near ($\pi $,$\pi $) \cite{10}), and metallic
islands (the latter probably being of a MFL character).

Numerous evidences currently fully confirm such picture for PG regime. For
instance, the motion of phases may be slow down by different defects which
results in the so-called ``Cu wipe-out'' effect observed in \cite{11,12,13},
provided frequencies of fluctuations get low enough to approach the NMR
frequency window. Gradual ``freezing'' of fluctuations and subsequent glassy
localization of heterogeneities has been demonstrated in \cite{13A}. Static
stripes are known for the Nd- and Eu- doped LSCO \cite{14,15,16} and in LCO
doped by barium, Ba, near the Ba concentration $x=1/8$ commensurate to the
periodicity of the low temperature tetragonal lattice phase \cite{17}.
Finally, the unequivocal confirmation in favor of two phase coexistence
comes from the very fact that at low enough temperature (usually below $T_{c}
$) the ICAF is seen in the neutron diffraction experiments \cite{18},
proving onset of their static coexistence in the real space at low
temperatures \cite{19,20} (fraction of the ICAF phase increases with applied
fields).

As it was mentioned, below we make an attempt to derive from the
experimental data indications whether the amount of the Sr- doped holes
determines the \textit{total} number of charge carriers in cuprates. Our
conclusion is that the number of carriers increases faster than $x$ at
higher $x$ and $T$. For the electronic spectrum of cuprates, it also
signifies that the Cu($d^{9}$)-levels mix together with the holes on the
oxygen ions into a common band that is studied by the ARPES experiments.

We considered the available experimental data on the Hall coefficient at
elevated (up to 1000K) temperatures \cite{21,22,24}. We have found that in a
broad range of $x$ and temperatures the data for carriers concentration from
the Hall measurements, $n_{Hall}$, can be presented surprisingly well in a
form
\begin{equation}
n_{Hall}=n_{0}(x)+n_{1}(x)\exp (-\Delta (x)/T)  \label{eq1}
\end{equation}%
(It turnes out that Eq. (1) also describes new results \cite{23}). The $x$%
-dependence of the temperature independent component, $n_{0}$, is given in
Fig.1. Aside from the scattering at small $x$ $n_{1}(x)$ is practically
\begin{figure}[tbp]
\centering \includegraphics[height=8cm]{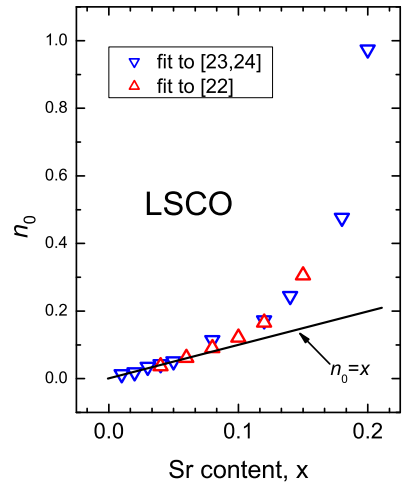} 
\caption{The doping dependence of $n_{o}(x)$, obtained by the fitting of
Eq.(1) to the experimental Hall coefficient temperature dependence %
\protect\cite{21,22,24} for LSCO.}
\end{figure}
constant ($\sim 2.8)$ up to the vicinity of $x=0.2$, where it drops down
abruptly. Note considerable deviations from the linear in $x$ behavior in $%
n_{0}$! Although numerous factors (anisotropy, the temperature dependent
scattering processes, etc.) can complicate the theoretical interpretation of
the Hall effect data, it is known, however, that interactions drop out for
the isotropic Fermi-liquid model \cite{25} and even for a non-parabolic but
isotropic shape of the FS \cite{26}. Meanwhile, it is seen in Fig.1 that
deviations from the linear dependence begin rather early, already at $x=0.07$%
. The hole-like FS ``locus'' seen by ARPES, being centered at $(\pi ,\pi )$,
is practically isotropic up to $x=0.11$ \cite{27,28}. Therefore even the
temperature independent amount of carriers grows faster than $x$ with doping
(similar results have been reported \cite{29} for Bi$_{2}$Sr$_{2-x}$La$_{x}$%
CuO$_{6}$).

Unlike the Hall effect data for $n_{0}(x)$ where common interpretation
becomes unreliable at larger $x$, the activation character of the
temperature-dependent term in Eq. (1), however, is the thermodynamic feature
and, as such, should not be sensitive to model scattering mechanisms. The
exponential contribution describes the thermally activated carriers that
come from levels lying deeply below the chemical potential.

In principle, such a term could come from different regions of the material
because of its inherent non-homogeneity in the real space. We assume,
however, that activated carriers are excited from some deeper parts of the
LSCO energy bands. Indeed, according to the ARPES data \cite{4, 28, 30}, in
addition to the ``coherent'' states corresponding to an ``arc'' on the locus
of FS at the chemical potential, in the vicinity of $(0,\pi )$ and other
symmetric points there are seen deeply-lying energy bands featured by the
high density of electron states (so called van Hove ``troughs'', or van Hove
singularities). We suggest that electrons are thermally activated from the
van Hove ``troughs'' at $(0,\pi )$ and the rest of the symmetric points, and
go into the vicinity of the nodal ``arc'' at the chemical potential on the
FS ``locus'' \cite{4,28}, where the activated electrons join the liquid of
mobile carriers. In order to check this suggestion we have plotted in Fig. 2
data for the energy gap $\Delta (x)$ of Eq.(1) together with the energy
separating the underlying van Hove bands from the Fermi level which was
deduced from Fig. 3 of \cite{4} and Fig. 3b in \cite{28} for various Sr
\begin{figure}[tbp]
\centering \includegraphics[height=8cm]{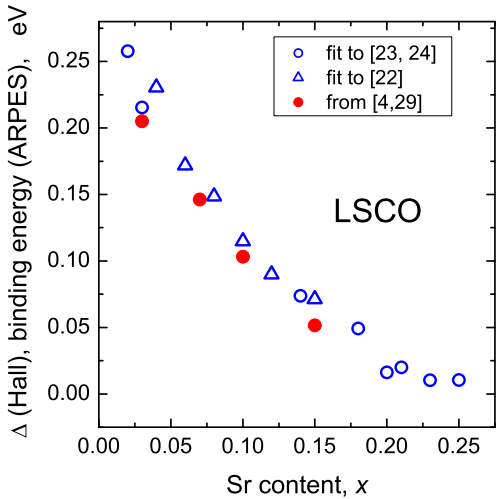} 
\caption{The activation energy $\Delta (x),$obtained by the fitting of Eq.1
to the experimental Hall coefficient temperature dependence \protect\cite%
{21,22,24} for LSCO (open circles and triangles), $vs$ the energy separating
the underlying van Hove bands from the Fermi level (binding energy) which
was deduced from Fig. 3b of \protect\cite{4} and Fig. 3 in \protect\cite{28}
for various Sr concentrations (shown with filled circles.}
\end{figure}
concentrations (shown with red circles). For both quantities the extracted
values are in an excellent agreement, thus giving the strong argument in
favor of our interpretation.

A mere band singularity in density of states at the van Hove point by
itself, of course, would be not enough to account for the thermodynamic
activation contribution. We argue that in ARPES one measures only one
component of the total energy of an electron imbedded into the lattice. In
other words, implicitly, we invoke a localization of electrons near the
vicinity of the van Hove points. Although from a different point of view,
importance of lattice/polaronic effects at the interpretation of ARPES data
has been pointed out also in \cite{Shen}. (We estimate the order of
magnitude for the lattice component as a few tens meV, see below).

Eq. (1) describes well the Hall data \cite{21,22,23} practically in the
whole available temperature interval without revealing sharp features or
changes in the behavior near the line of the hypothetical 1st order
transition, $T^{\ast }(x)$, mentioned above. It is worth to emphasize that
there are no reasons for appearance of such features at the onset of the
transition, because the transition does not realize itself: the frustrations
caused by the Coulomb forces allow only fluctuations corresponding to a
dynamical two-phase coexistence in the PG regime, instead of the macroscopic
phase segregation. Consequently, $T^{\ast }(x)$ marks only a crossover
between the left- and right hand sides of the $(T,x)$-phase diagram for LSCO
cuprates. A good fit for $T^{\ast }(x)$ is obtained just from the comparison
when the number of doped carriers, $x$, and activated ones become
approximately equal:
\begin{equation}
T^{\ast }(x)\approx T_{0}(x)=-\Delta (x)/\ln x  \label{eq2}
\end{equation}%
In Fig. 3 we plotted $T^{\ast }(x)$ defined according to Eq.(2)
and the crossover temperatures obtained differently from other
experiments.

The fact that the decomposition into two contributions given by Eq.(1)
covers the PG regime at smaller temperatures \cite{22} and even reproduces
well the low temperatures Hall measurements in the high fields normal state
of LSCO \cite{31} raises questions. In fact, to which of the two PG
sub-phases one is to correlate the Hall effect data? The same question
concerns the ARPES measurements.

It seems reasonable to connect the Hall coefficient, $R_{H}$, with
properties of the metallic component. Indeed, if stripes were pinned by
\begin{figure}[tbp]
\centering \includegraphics[height=8cm]{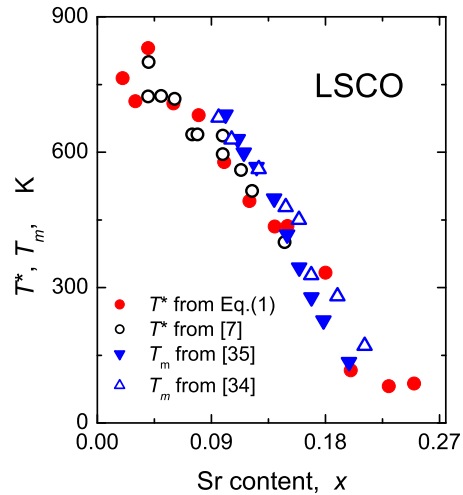} 
\caption{The psedogap crossover temperature $T^{\ast }(x)$ obtained: with
the help of Eq.(1) with the activation energy $\Delta (x)$ shown in Fig. 2
(filled circles); from the crossover temperature of the resistivity curves
(open circles); as the temperature $(T_{m})$ corresponding to the maximum of
the magnetic susceptibility measured in \protect\cite{31A} ( filled
triangles) and \protect\cite{31B}(open triangles).}
\end{figure}
defects, and the conductivity along the stripes were expected to bear the
one-dimensional character, that should suppress conduction in the transverse
direction. These arguments have been tested \cite{32} for the temperature
dependence of the Hall coefficient in the Nd-doped LSCO material with
different Sr content. The Hall coefficient, $R_{H}(T,x)$ \cite{32} has a
characteristic drop at temperatures $T\sim 70-100K$, at least for
under-doped compositions. One finds similar features in the data \cite{31}
that even are characterized by approximately the same temperature scale. It
is important that this effect is strongest for $x=0.12$, where the fraction
of the stripe phase should be maximal.

As to the second question, since the ARPES experiment is a fast measurement,
it provides the instant snapshot of deep energy levels for metallic islands
and probably does not change essentially when taken above or below $T^{\ast
}(x)$ \cite{32A}. Making use of the excellent agreement in Fig.2 of the Hall
activation energy with the ARPES results \cite{4,28}, we have extended our
gap analysis in terms of Eq.(1) to the data \cite{22} at higher $x$.
Notable, the result shows a plateau at $x$ just above $\sim 0.2$ in Fig.2.
According to \cite{4,27,28}, the FS ``locus'' experiences the topological
change from the hole-like ``FS'' centered at $(\pi ,\pi )$ to the
electron-like one centered at $(0,0)$ very close to $x=0.2$. ``Gap'' seen in
Fig.2, obtained from the interpolation of Eq.(1) into this concentration
range, in our opinion produces the estimate for characteristic energy scales
of the lattice effects ($\sim 10$meV).

As it is known, in the ARPES experiments one start seeing the FS ``locus''
in LSCO already at concentration as small as $x=0.03$. However the ``FS''
obtained in this way covers a large area of Brillouine zone that then
changes in ``agreement'' with the ``Luttinger count'', $1-x\ $\cite{27}.
Since currently it is established \cite{4} that the propagating (coherent)
excitations come about only in a narrow ``arc'' near the nodal directions,
there are no contradiction between \cite{27} and the trend seen in Fig.1
that shows that actual number of mobile carriers grows faster than $x$ even
at lower temperatures.

Since we have already touched above some issues related to resistivity of
cuprates, it is worth mentioning a peculiar feature that, in a sense, is
fully consistent with our general line of arguing. For extremely small Sr
doping the thermal excitation of carriers gives rise to very interesting
transport behavior - the temperature independent contribution to
conductivity. It comes about due to the fact mentioned above that each
thermally activated charge creates a local defect at the CuO$_{2}$ plane.
These defects play the role of scattering centers and contribute to
resistivity at high temperatures. The density of these defects equals to the
density of charge carriers produced by thermal activation. Hence the same
activation energy governs the lifetime of charge carriers. It results in
temperature independent resisitvity. This property manifests itself
experimentally as saturation of the temperature dependence for resistivity
at extremely low doping \cite{24}.

Contributions to the activated component of the Hall coefficient in Eq.(1)
come from the vicinity of the van Hove bands that have a pronounced 1D
behavior. Therefore the emptied sites should reveal a localized behavior. We
suggest that activated carriers in Eq.(1) may add a temperature independent
contribution into resistivity at high temperatures as well.

A simple consideration that does not compromise the ideas of the MFL \cite{8}%
, but may still be essential for understanding of the linear in T
resistivity \textit{well below }$T^{\ast }$ (e.g. for such doping level as $%
x=0.15$ \cite{33,34}, is that deeply inelastic scattering processes, by
removing one of the conservation laws' constraints in the ordinary FL
approach, would immediately produce such linear dependence.

Finally, we note by passing the specific features in the behavior $n_{0}(x)$
in Fig.1 near $x\sim 0.2$ (we mentioned above the drop of $n_{1}(x)$ at the
same $x$). This concentration has already been identified in a number of
publications as an emerging QCP for cuprates \cite{35,36}. Appearance of the
plateau in Fig. 3 at exactly the same concentration is in the agreement with
this expectation.

To summarize, we found the quantitative agreement between the activation
energies in the high temperatures Hall data and the ARPES measurements. It
also has been shown that the actual concentration of mobile carriers is not
equal to the number of the externally introduced holes, even more, the
carriers concentration increases with doping and temperatures. In turn, it
signifies that Cu spins are not fixed at a given Cu-site. In other words,
spins may move along, in agreement with other arguments \cite{9} that
consider PG region as a region of dynamically coexisting and competing
sub-phases.

\begin{acknowledgements}

The authors are grateful to Y. Ando for sharing with us the recent
Hall measurements data \cite{23} prior to publication. We also
appreciate helpful discussion with D. Pines. The work of L.P.G.
was supported by the NHMFL through NSF cooperative agreement
DMR-9527035 and the State of Florida, that of  G.B.T. through the
RFBR Grant N 04-02-17137.
\end{acknowledgements}

\bigskip


\begin{references}
\bibitem{32A} compare with \cite{24}: based on the analysis of their optical data the authors [24] came to the conclusion that "the local environment of mobile holes remains unaltered by doping"
\end{references}
\end{document}